\documentstyle[aps,epsf,rotate,multicol,amssymb]{revtex}
\begin{document}


\title{First-order  phase transitions in non-equilibrium
systems: New perspectives}
\author{Roberto A. Monetti, Alejandro Rozenfeld, and Ezequiel V. Albano}
\address{Instituto de Investigaciones Fisicoqu\'{\i}micas Te\'oricas y 
Aplicadas (INIFTA), UNLP, CONICET, CIC (Bs. As.), C. C. 16 Suc. 4,
1900 La Plata, Argentina} 
\date{\today}

\maketitle

\begin{abstract}
First-order irreversible phase transitions (IPT's) between an active regime and an
absorbing state are studied in two models by means of both 
simulations and mean-field stability analysis. Hysteresis around
coexistence is the result of the interplay of the length of the
interface, its curvature and a memory effect related to the phase
which is being removed. A controversy on the occurrence of scale-invariance
is clarified, conciliating the behavior of IPT's with its
reversible counterpart.
\end{abstract}

Pacs numbers: 05.70.Jk, 64.60.Ht, 02.50.Ey, 82.65.Jv

\begin{multicols}{2}

The study of far from equilibrium systems continues to attract great attention
and has become challenging subject of interest for many areas of research in
physics, chemistry, ecology, catalysis, economy, social sciences, 
etc. \cite{sci,ron,eze}. An intriguing feature of those systems is the
occurrence of irreversible phase 
transitions (IPT's) between an active regime and an  absorbing state
where the system becomes trapped.
After the work of Ziff et. al. \cite{ZGB}, our understanding of
second-order IPT's has experienced a rapid growth since they have
unambiguously been placed either in the directed percolation or parity 
conserving universality classes \cite{eze,roberto}. Due to the lack of experimental
feedback on second order IPT's, the huge activity in 
the field is mainly of academic-theoretical interest.
In spite of the existence
of experimental evidence on systems undergoing first order IPT's \cite{ehs},
our understanding of this field is far from being satisfactory. There
are important controversies and many 
aspects still remain unexplored. 
In fact, the claim that 
power-law behavior could also hold for first order IPT's is certainly a
puzzle \cite{jim}. A similar controversy has recently raised in the
field of reversible transitions \cite{gul}. However, power law behavior can be identified
as finite-size effects which vanishes in the thermodynamic
limit \cite{gul}. Furthermore, the existence of hysteresis, which is a signature
of first order transitions in equilibrium systems, has so far, not
been explored in detail in the field of IPT's. 

The aim of this
work is to present an study of first order IPT's based on two complementary
techniques, namely extensive numerical simulations and mean field approaches.
Two different model systems, namely the ZGB model
for a catalyzed reaction \cite{eze,ZGB} and stochastic game of life (SGL) for a
society of living individuals \cite{gl} are investigated in order to
allow the discussion of useful comparisons.

The SGL model is a cellular automaton defined on
a square lattice where each site $\sigma_{ij}$ can take only two 
values $\sigma_{ij} = \{0 \; \text{(dead site)},1 \; \text{(living site)} \}$
and interacts with its eight nearest neighbors which defines its
neighborhood (NH).  
The system evolves from a given time to the next time step as follows:
i) A living site whose NH is dead or allocates one living
site, will die. 
ii) A living site whose NH allocates more than three living sites,
will die. 
iii) A living site whose NH allocates two or three living sites, 
will survive with a probability $p_s$. 
iv) A dead site whose NH is dead or allocates one living
site, will remain in this state. 
v) A dead site whose NH allocates more than three living sites,
will remain in this state. 
vi) A dead site whose NH allocates two living sites, 
will become a living site with a probability $p_b$. 
vii) A dead site whose NH allocates three living sites, 
will become a living site. 

The phase diagram of the SGL 
has two phases,
namely extinction and life, both separated by a first order coexistence curve
\cite{gl}. A mean field (MF) equation for the time evolution of
the density of living sites $x$ is given by

\begin{eqnarray}
\frac{d x}{dt} = x[-x^{8} - 8 x^{7}y-28x^{6}y^{2} 
- 56x^{5}y^{3} - 
70x^{4}y^{4}  \\ -56(1-p_{s})x^{3}y^{5} 
+ 28(1+p_{s})x^{2}y^{6} 
+ (28p_{b} - 8)xy^{7} - y^{8}] \nonumber
\label{eq1}
\end{eqnarray}
where $x$ is the order parameter and $y = 1 - x$. 
The fixed points of eq. (1) satisfy $\frac{dx}{dt} = f(x) = 0$, and
the stable ones correspond to the stationary states of
the system. We define a potential $V(x)$ through the relation
$\frac{dx}{dt} = f(x) = - \frac{d V(x)}{dx}$. Then, the stable
(unstable) fixed points correspond to the minima (maxima) of $V(x)$,
respectively. The values ($p_s^{coex}, p_b^{coex}$) that satisfy 
$\frac{dV(x)}{dx} = 0$ and $\frac{d^{2}V(x)}{dx^{2}} = 0$ determine the MF
coexistence curve. Figure 1 shows a plot of $V(x)$ as a function of $x$ and 
$p_b$, keeping $p_s = 0.1$ fixed. $x = 0$ is 
{\bf always} a minimum of $V(x)$ which corresponds to the absorbing state. In
addition, the surface exhibits another valley at a higher density
which corresponds to the stationary living phase.
The valley at a higher density is vanishing when
decreasing $p_b$, and it 
actually disappears at the coexistence point $p_b^{coex}$ given by
($p_{s} = 0.1, p_b^{coex} = 0.3131$) (see figure 1).
Further decreasing $p_b$ leads to 
potential functions displaying only the valley at $x=0$. Figure 1 clearly 
shows that the valley at a higher density disappears at a value $x_c$ well
above $x=0$. So, a sharp jump  in the order parameter 
of the system is observed indicating a first order IPT 
in agreement with the simulation
results \cite{gl}. Since the system always evolves to the potential
valleys, the stationary state will depend
upon the initial density. This MF result is also in agreement
with simulations.

The ZGB model is an approach to the catalytic oxidation of carbon monoxide,
$CO + 1/2 O_2 \rightarrow CO_2$ \cite{ZGB}.
The arrival probabilities of $CO$ and $O_2$ are normalized,
$P_{CO}+P_{O_2}=1$, so one has a single parameter which is selected to
be $P_{CO}$. The ZGB model exhibits a first order IPT between a regime
of sustained reaction and a CO-poisoned state close to $P_{2CO}=
0.5256(1)$ \cite{nota}. For details on the ZGB model see e.g. \cite{eze,ZGB,ZB}.
\begin{figure}
\narrowtext
\centerline{{\epsfysize=2.4in \epsffile{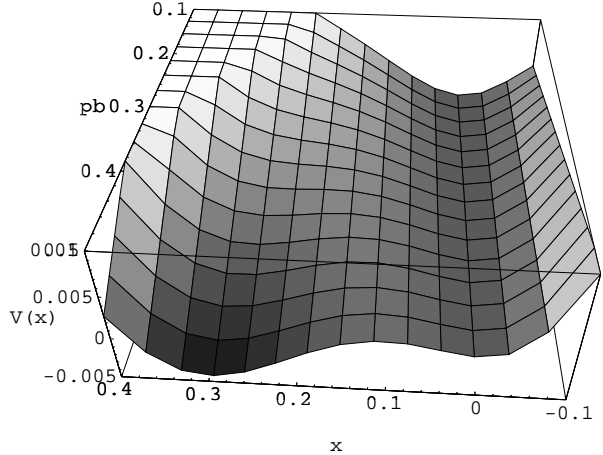}}}
\caption{3d plot of the potential $V(x)$ for $p_s = 0.10$,
corresponding to the SGL model. Darkest areas correspond to the potential minima.}
\label{fig1}
\end{figure}

In this paper, hysteresis in the SGL model has been studied using the
constant-coverage (CC) ensemble \cite{ZB}. First, one has to achieve a
stationary configuration using the standard ensemble. Then, the system
is switched to the CC ensemble where the density $x$ of living sites is varied
stepwise. Since the SGL model has two external parameters, $p_s$ and 
$p_b$, in the CC ensemble $p_s$ is kept constant and the control parameter is now $x$. 
Each point of the phase diagram, $x$ versus $p_b$ (see figure 2), is
evaluated after allowing the system to relax $\tau_{R}$ time steps. The
phase diagram is identical to that obtained with the standard method
only within the living phase, i. e. for $p_b > p_b^{coex}$. 
When approaching the coexistence point, the curve bends downwards,
close to $p_b^{coex}$, becoming parallel to the $x$ axis. Then, the value of the
parameter $p_b$ remains constant, within this density range, pointing out 
that the living and the absorbing phase coexist. 
The MF phase diagram ($x$ versus $p_b$) exhibits two branches of fixed
points, the upper stable curve and the lower unstable one, that
intersect at the coexistence point where the sharp jump in the density occurs. 
Further decreasing the parameter $p_b$, leads to only one stable 
stationary state, namely the absorbing state. Then, no stable
stationary living state exists for densities $0 < x < x_c$. Using the 
the CC ensemble the system is forced to support a density within this
range, so it splits in two stable phases, namely the living phase at
$p_b = p_{b}^{coex}$ and the absorbing phase that behave almost 
independently. This is the reason why $p_b$ remains constant in the
coexisting region. 
After achieving a density close to the
absorbing state, we begin to slowly increase it. At that stage, a big
absorbing cluster has occupied most of the system. Since the density
can only grow from the interface between the coexisting phases by
invading the absorbing cluster, a larger value of $p_b$ is
needed. This is observed in figure 2, where the increasing density
branch is the rightmost branch of the loops. 
\begin{figure}
\narrowtext
\centerline{{\epsfysize=2.4in \epsffile{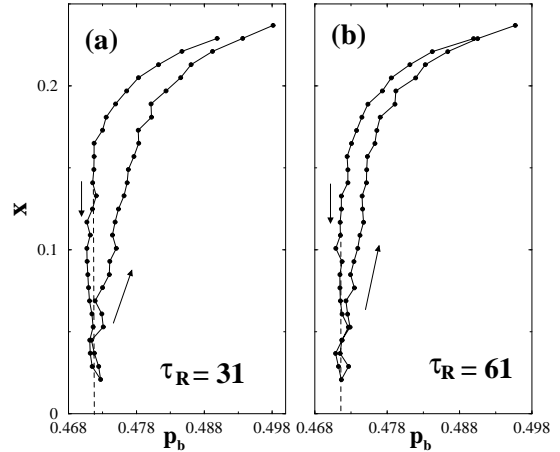}}}
\caption{Hysteresis loops obtained by means of the CC ensemble for the
SGL model, keeping the parameter $p_s = 0.1$ fixed, for different
relaxation times $\tau_R$. Loops are generated counterclockwise (see
arrows). The dashed line indicate the position of the
coexistence point.}
\label{fig2}
\end{figure}
This invading process,
which can be regarded as a memory effect related to the absorbing
cluster, is unique for the growing branch. It is also found that the
longer the relaxation time $\tau_{R}$ the narrower the loops. In
addition, the left branch (decreasing density branch) seems to fall
downwards at the same $p_b$ value independent of the value of $\tau_{R}$. Then, the
right branch approaches the left one for longer $\tau_{R}$. After a
view inspection of two configurations obtained at the same density but
on different branches (not shown here),
it is clear that the presence of a large absorbing cluster in
the configuration corresponding to the right branch and the existence
of many small clusters in the ones corresponding to the left branch is the main
difference among them. Then, the length of the interface between the
two phases is clearly larger for configurations on the left branch,
and consequently they are more efficient at keeping
the population constant. This results in a
lower value of the parameter $p_b$. Notice that along the decreasing
branch the memory effect is due to the presence of small active
clusters. Also, loops become narrower
for longer $\tau_{R}$ because the system evolves towards the most disordered state
with the largest interface at a given density.

In the case of the ZGB model, the density of  $CO$
($\theta_{CO}$) plays the role of the control parameter (figure 3). 
Hysteresis effects are absent in small lattices
($L \leq 64$), while hysteresis loops can be 
distinguished for larger lattices
(figure 3). So,
for the CO-growing (decreasing) branches of the loops we obtain
$P_{CO}^{G}\cong 0.52641(1)$ ($P_{CO}^{D}\cong 0.52467(3)$), respectively.
Notice that the value of the parameter remains almost constant
for a wide range of $\theta_{CO}$ values ($0.1\leq\theta_{CO}\leq 0.90$)
for $L\geq 512$.
\begin{figure}
\narrowtext
\centerline{{\epsfysize=2.4in \epsffile{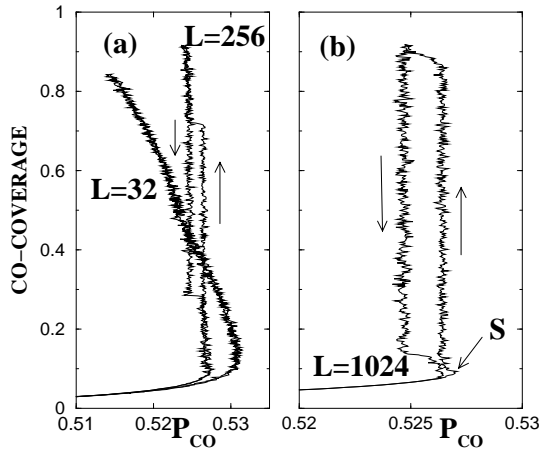}}}
\caption{Plots of $\theta_{CO}$ versus $P_{CO}$ obtained using the CC
ensemble with $\tau_R=100$ mcs and lattices of different size. The
point S shows the position of the upper spinodal point
$P_{CO}^S$. Arrows pointing up and down show the growing and
decreasing $\theta_{CO}$ branches of the hysteresis loop, respectively.}
\label{fig3}
\end{figure}
Assuming that the coexistence point lies
in the middle of the loop, we get $P_{2CO} = 0.52554(4)$. We claim
that this value, obtained by measuring 400 different points each of them 
averaged over $2x10^3$ Monte Carlo steps, is the most accurate available \cite{nota}.
Snapshot configurations were obtained for both branches (not shown here).
For the growing branch a massive $CO-$cluster preveals and after
becoming large enough, it percolates along only 
one direction of the lattice, forming two (relatively flat) interfaces at the
boundary of the coexisting phases. When $\theta_{CO}$ is increased the interface
roughens and eventually dangling ends evolving from both parallel
interfaces get in contact causing the onset of percolation in both directions of
the lattice. At this point the system jumps to the other branch of the
hysteresis loop. So, the interface between coexisting phases has
become shorter and its curvature has decreased. Smaller local
curvature leads to a less efficient $O_2$ adsorption mechanism and
consequently a larger $O_2$ pressure (smaller $P_{CO}$) is needed in
order to decrease the $\theta_{CO}$. In addition, the memory effect
associated to a bulky active phase, results in a bigger $P_{CO}$ value
along the growing branch, in contrast to the memory effect along the
decreasing branch which is related to a bulky $CO$ cluster.
The phase diagram of the ZGB model as obtained by the CC-ensemble, 
clearly exhibits the upper-spinodal point $P_{CO}^S$, which 
strongly depends on the lattice size. In the infinite-size limit
($L=\infty$), we obtain the value 
$P_{CO}^S(L=\infty) = 0.5270(5)$ which may be compared with
$P_{CO}^S(L=256) \simeq 0.527$ \cite{ZB} and
$P_{CO}^S(L=?) \simeq 0.5285$ \cite{jim}. An stability analysis, similar to that 
described above for the SGL, has also been performed within the MF
approach proposed by Dickman \cite{ron}. Both models display two
different branches of fixed points, namely a stable branch and an
unstable one, which coalesce at a point identified as the spinodal
point \cite{ron}. CC simulations clearly show the existence of a
spinodal point for the ZGB model but no evidence is found in the case
of the SGL model.

Another approach for the study of IPT's is the epidemic
analysis (EA) \cite{jim}. For the SGL, we have studied the time 
evolution of the average number of living sites ($N(t)$) as it is
shown in figure 4. Initializing
the simulation with a small colony of living sites in an otherwise
extinct state, 
it is found that the asymptotic regime is reached after $t > 10^4$ updates and 
it is possible to identify subcritical, critical and supercritical
curves as well. The critical point ($p_s=0.1$, $p_b=0.47188$) is in
excellent agreement with the coexistence point obtained by means of
the CC ensemble (see figure 2).
Notice the high sensitivity of $N(t)$ to tiny changes in the values
of the parameter. $N(t)$ also displays a short time
regime ($t < 10^2$) and an intermediate time regime ($10^2 < t <
10^4$), the last resembling a plateau behavior. 
\begin{figure}
\narrowtext
\centerline{{\epsfysize=2.4in \epsffile{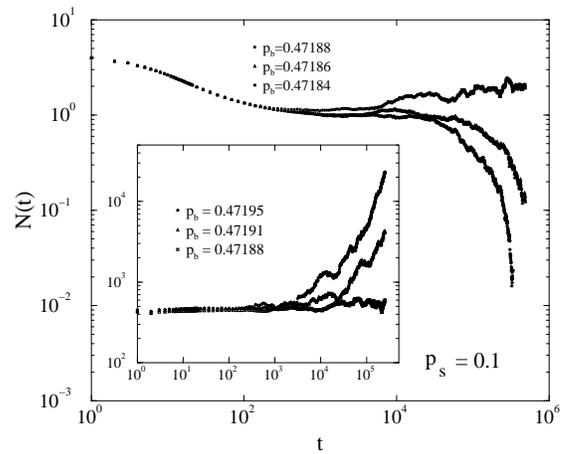}}}
\caption{Log-log plots of the average number of living sites $N(t)$
versus $t$ for different values of the parameter $p_b$, keeping $p_s =
0.1$ fixed. The EA are initialized using small living colonies (bigger
colonies) for the main plot (inset plot), respectively.}
\label{fig4}
\end{figure}
Starting with bigger colonies (inset of figure 4), we observe the
following three main differences, i) the short time behavior is absent, 
ii) intermediate time regime dominates from the very beginning and,
iii) asymptotic regime is achieved sooner ($t \approx 10^3$).
These features
can be understood on the basis of the
potential $V(x)$. In fact, Figure 1 shows that for low initial
densities, $x$ will flow to $x(t \rightarrow \infty)=0$. This is  
the reason why $N(t)$ displays a decreasing short time behavior in
figure 4, which appears to be a universal feature
observed in every EA of a first-order IPT.  
However, in some few cases and due to a density fluctuation, the
system may overcome the  potential barrier, reaching a small region
where $V(x) \approx constant$. 
Consequently, the system can remain for a long time in a region where 
$x \approx constant$ until another fluctuation would drive it either towards $x=0$
or to the stationary living state. This explains the plateau observed
in the intermediate time regime and the asymptotic regime as well.
Another way of corroborating the above explanation is to initialize the
simulation with larger colonies (see inset of figure 4). So, as the density is in a
region where $V(x) \approx constant$ from the very beginning, 
the short time behavior is absent and only the plateau and the
asymptotic time behavior remain.The short time behavior of the system
can be easily confused with a power-law behavior, mainly if small
lattices and short times are considered.

In the case of the ZGB model the absorbing state is a lattice
fully covered by $CO$. So, the EA
is started with a covered lattice except by a small amount of empty sites
placed at the center of the sample. Figure 5 shows results for the
number of empty sites ($N(t)$) versus the  time $t$. 

\begin{figure}
\narrowtext
\centerline{{\epsfysize=2.5in \epsffile{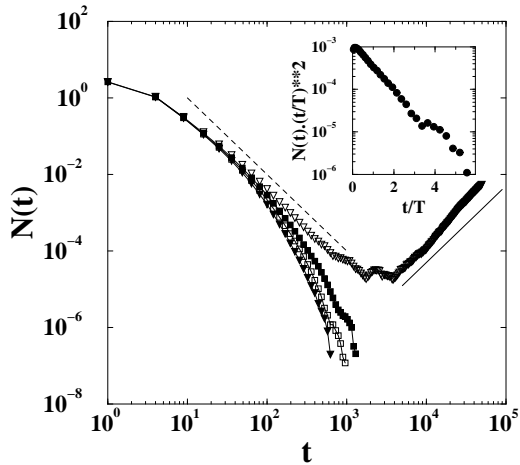}}}
\caption{Log-Log plots of the number of vacant sites $N(t)$ versus $t$ for
EA of the ZGB model. Results averaged over $5 x 10^8$ different runs
($\blacktriangledown$ $P_{CO}^{G}$, $\blacksquare$ $P_{CO}^{D}$, $\square$
$P_{2CO}$, $\bigtriangledown$ $P_{CO} = 0.52345$).
For the later, two straight lines have been drawn for the sake of
comparison: the dashed one with $\eta^{eff}=2$ and the full one with
slope $2$, respectively. The inset shows a semi-logarithmic plot of
$N(t)(T/t)^{-2}$ versus $t/T$ with $T=183$, according to equation 2.}
\label{fig5}
\end{figure}
Results obtained for $P_{CO} \leq P_{CO}^{G}$
show pronounced curvature, with clear evidences of a cut-off. So, the
dynamical critical behavior of the ZGB model at coexistence does not
exhibit scale-invariance. This finding is in contrast with previous
results claiming power-law and scaling behavior \cite{jim}. For
$P_{CO} \gtrsim P_{CO}^{G}$, $N(t)$ exhibits a pseudo-power-law behavior over many 
decades with an effective exponent $\eta_{eff}\cong 2.0 \pm
0.1$. However, eventually after a long time, a successful epidemic
spreading preveals and $N(t)$ suddenly growths as
$N(t)\propto t^2$, as it is shown in figure 5, indicating a spatially
homogeneous spreading. In order to understand the critical behavior 
of $N(t)$ at coexistence we propose the following anzats

\begin{equation}
N(t)\sim (t/T)^{-\eta_{eff}} \exp (-t/ T),
\label{eq2}
\end{equation}
where $T$ sets a characteristic time scale which is validated by the plot in the
inset of figure 5. A regression analysis gives $T \cong 183 \pm 3$.
Summing up, one observes a pseudo power-law behavior for short time ($t < T$)
which crosses over to an asymptotic exponential decay for larger time. 

In summary, we have studied two different models that
exhibit first-order IPT's, using extensive computer simulations
and MF approximations. Despite of the lack of universality, the
critical behavior of both models share several features, namely (i)
hysteresis is absent for small lattices, (ii) hysteresis effects are
due to the interplay between differences in length and curvature of the
interfaces among the coexisting phases at the same density, and the existence of memory
effects related to the phase which is being removed. Which of the
three effects play a major role may depend on the system and is the
subject of an ongoing investigation,  
and (iii) the occurrence of power-laws in the dynamical
critical behavior of first-order IPT's can be safely ruled out. This
last finding conciliates the behavior of first order IPT's with their counterpart
in equilibrium systems where it is well established that the existence
of short range correlations inhibits the observation of scale
invariance. However, major differences to be noted are: (i) in
contrast to the ZGB model, spinodal points are not observed in the SGL
model, (ii) for $\tau_R \rightarrow \infty$, hysteresis loops collapse
to a single curve. However, for the SGL model there is a single
vertical line at coexistence, while for the ZGB model the curve splits
out in two vertical lines resembling a phase transition in the parameter
space, and (iii) the intermediate time dynamic critical behavior is different. 

{\bf Acknowledgments}: This work was supported by CONICET, UNLP, ANPCyT, 
Fundaci\'on Antorchas (Argentina), and Volkswagen Foundation (Germany).

\newpage



\end{multicols}

\begin{references}
\bibitem{nota} The error in the value $P_{2CO} = 0.52560(1)$
\cite{ZB}, seems to be optimistic on view of the size of the
lattices used.
\end{references}
\end{document}